\documentclass[twocolumn,showpacs,preprintnumbers,superscriptaddress,prl,amsmath,amssymb]{revtex4}

\usepackage{graphicx}
\usepackage{dcolumn}
\usepackage{bm}

\begin{document}

\title{Flux growth and physical properties of Mo$_3$Sb$_7$ single crystals}

\author{J.-Q. Yan}
\affiliation{Materials Science and Technology Division, Oak Ridge National Laboratory, Oak Ridge, TN 37831, USA}
\affiliation{Department of Materials Science and Engineering, University of Tennessee, Knoxville, TN 37996, USA}

\author{M. A. McGuire}
\affiliation{Materials Science and Technology Division, Oak Ridge National Laboratory, Oak Ridge, TN 37831, USA}

\author{A. F. May}
\affiliation{Materials Science and Technology Division, Oak Ridge National Laboratory, Oak Ridge, TN 37831, USA}

\author{H. Cao}
\affiliation{Quantum Condensed Matter Division, Oak Ridge National Laboratory, Oak Ridge, TN 37831, USA}

\author{A. D. Christianson}
\affiliation{Quantum Condensed Matter Division, Oak Ridge National Laboratory, Oak Ridge, TN 37831, USA}

\author{D. G. Mandrus}
\affiliation{Materials Science and Technology Division, Oak Ridge National Laboratory, Oak Ridge, TN 37831, USA}
\affiliation{Department of Materials Science and Engineering, University of Tennessee, Knoxville, TN 37996, USA}

\author{B. C. Sales}
\affiliation{Materials Science and Technology Division, Oak Ridge National Laboratory, Oak Ridge, TN 37831, USA}

\date{\today}
\begin{abstract}
Millimeter sized single crystals of Mo$_3$Sb$_7$ are grown using the self-flux technique and a thorough characterization of their structural, magnetic, thermal and transport properties is reported. The structure parameters for the high-temperature cubic phase and the low-temperature tetragonal phase were, for the first time, determined with neutron single crystal diffraction. Both X-ray powder diffraction and neutron single crystal diffraction at room temperature confirmed that Mo$_3$Sb$_7$ crystallizes in Ir$_3$Ge$_7$-type cubic structure with space group \emph{Im}$\overline{3}$\emph{m}. The cubic-tetragonal structure transition at 53\,K is verified by the peak splitting of (4\,0\,0) reflection observed by X-ray single crystal diffraction and the dramatic intensity change of (12\,0\,0) peak observed by neutron single crystal diffraction. The structural transition is accompanied by a sharp drop in magnetic susceptibility, electrical resistivity, and thermopower while cooling. A weak lambda anomaly was also observed around 53\,K in the temperature dependence of specific heat and the entropy change across the transition is estimated to be 1.80\,J/molMo\,K. The temperature dependence of magnetic susceptibility was measured up to 750\,K and it follows a Curie-Weiss behavior above room temperature. Analysis of the low-temperature magnetic susceptibility suggests a spin gap of 110\,K around 53\,K. A typical phonon thermal conductivity was observed in the low temperature tetragonal phase. A glassy phonon thermal conductivity above 53\,K suggests a structural instability in a wide temperature range. Superconductivity was observed at 2.35\,K in the as-grown crystals and the dimensionless specific heat jump $\triangle$C(T)/$\gamma$$_n$T$_c$ was determined to be 1.49, which is slightly larger than the BCS value of 1.43 for the weak-coupling limit.

\end{abstract}

\maketitle

\section{Introduction}
Despite a relatively low superconducting transition temperature T$_c$ = 2.08\,K,\cite{Bukowski2002SSC} the Zintl compound Mo$_3$Sb$_7$ has attracted considerable interest due to the possible involvement of magnetism in superconducting pairing,\cite{Candolfi2007PRL,Tran2008PRL,Tran2009JP,Oles2008} and promising thermoelectric performance with proper doping.\cite{Parker2011,HTTEMo3SbTe,HTTERu,Jeff2011} Mo$_3$Sb$_7$ is the only known compound in the Mo-Sb binary system. It crystallizes in a Ir$_3$Ge$_7$-type cubic structure with space group \emph{Im}$\overline{3}$\emph{m} at room temperature. The Mo-sublattice is characterized by a three dimension network of Mo-Mo dumbbells formed by nearest neighbours (NN) and octahedral cages at the body center positions formed by next nearest neighbours (NNN). The competition between strong antiferromagnetic NN and NNN interactions is proposed to drive a structure transition at T*\,=\,53\,K from cubic to tetragonal (space group \emph{I}4/\emph{mmm}) to relieve spin frustration.\cite{ValenceBondXtal,StructureJP,StructureMRB} This structure transition further shortens the NN Mo-Mo distance along the crystallographic c-axis and forms spin singlet dimers. $\mu$SR, inelastic neutron scattering, and magnetic susceptibility measurements on polycrystalline samples suggest that the formation of a 120\,K spin gap accompanies the structural transition upon cooling.\cite{Tran2009JP,Tran2008PRL} With a structure transition and a spin gap above T$_c$, Mo$_3$Sb$_7$ is a unique system for exploring the interplay between magnetism, structure instability, and superconductivity.

Despite much effort, the pairing mechanism at low temperatures and the correlation between the structural transition, magnetic frustration, and superconductivity are still controversial. Thorough studies of the intrinsic properties of Mo$_3$Sb$_7$ and related compounds have been hampered by the unavailability of sizable single crystals. The first attempt to grow Mo$_3$Sb$_7$ single crystals was performed by chemical transport reactions and resulted in small crystals adequate for determining the crystal structure.\cite{Jensen1966} Millimeter sized single crystals suitable for some physical property measurements were first grown by melting Sb in a thick Mo tube.\cite{Bukowski2002SSC} Upon cooling from 1000$^o$C to 700$^o$C at the rate of 3$^o$C/h, single crystals grow on the walls of the Mo tube. Crystals grown in this manner become superconducting at T$_c$ = 2.08\,K, and enabled several studies of the intrinsic properties.\cite{Bukowski2002SSC,Dmitriev2007,Tran2008JOAM} However, these crystals are still small in
size. More importantly, it's difficult to grow crystals doped with other elements, especially substitutions at the Mo site.

In this paper, we report the flux growth of millimeter sized Mo$_3$Sb$_7$ single crystals and a thorough characterization of the crystals by x-ray powder diffraction, x-ray and neutron single crystal diffraction, magnetic susceptibility, electrical resistivity, thermoelectric power, thermal conductivity, and specific heat measurements. Crystals grown by the self-flux technique posses the highest T$_c$ reported to date and are free of magnetic impurities. Neutron single crystal diffraction was performed, for the first time, to determine the structure parameters of the cubic phase at room temperature and the tetragonal phase at 4\,K. The structural transition is accompanied with anomalies in the temperature dependence of electrical resistivity, magnetization, specific heat, thermopower and thermal conductivity.

\section{Experimental Details}

Single crystals of Mo$_3$Sb$_7$ were grown out of Sb flux. The starting materials are elemental
Mo powder (Alfa, 99.999\%) and Sb (Alfa, 99.9999\%). Mo powder was first reduced in flowing Ar balanced with 4\% H$_2$ for 12 hours at 1000$^o$C. Reduced Mo powder and commercial Sb shot were mixed in the ratio of Mo$:$Sb = 1$:$49 and placed in a 2\,ml Al$_2$O$_3$ crucible. A catch crucible of the same size containing quartz wool was mounted on top of the growth crucible and both were sealed in a silica ampoule under approximately 1/3 atmosphere of argon gas. The sealed ampoule was heated to 1000$^o$C in 5 hours and homogeneized for 12 hours before cooling to 700$^o$C over 100 hours in a programmable box furnace. At 700$^o$C, the Sb flux was decanted from the Mo$_3$Sb$_7$ crystals.

Elemental analysis of the crystals was performed using a Hitachi TM-3000 tabletop electron microscope equipped with a Bruker Quantax 70 energy dispersive x-ray system. X-ray diffraction from oriented single crystals and from powders ground from crystals was performed on PANalytical X'Pert Pro MPD powder x-ray diffractometer using Cu K$_{\alpha1}$. The room temperature x-ray powder diffraction pattern was refined by the Rietveld method using Fullprof.\cite{FullProf} Single crystal neutron diffraction was measured at HB-3A four-circle diffractometer at the High Flux Isotope Reactor at the Oak Ridge National Laboratory. A neutron wavelength of 1.003\,{\AA} was used with a bent perfect Si-331 monochromator.\cite{HB3A} The data were refined by the Rietveld method using Fullprof. Magnetic properties were measured with a Quantum Design (QD) Magnetic Properties Measurement System (MPMS) in the temperature interval 1.8\,K\,$\leq$\,T\,$\leq$\,750\,K. The temperature dependent specific heat and electrical transport data were collected using a 14 Tesla QD Physical Properties Measurement System (PPMS) in the temperature range of 1.9\,K\,$\leq$\,T\,$\leq$\,300\,K. Thermal conductivity and thermopower were measured from 1.9\,K to 300\,K using the Thermal Transport Option (TTO) from QD and a 9 Tesla PPMS. A rectagular bar with the dimension of 0.6\,mm\,$\times$\,0.7\,mm\,$\times$\,6\,mm was cut from a large crystal and used for the TTO measurement. Silver epoxy (H20E Epo-Tek) was utilized to provide mechanical and thermal contacts during the thermal transport measurements.

\begin{figure} \centering \includegraphics [width = 0.47\textwidth] {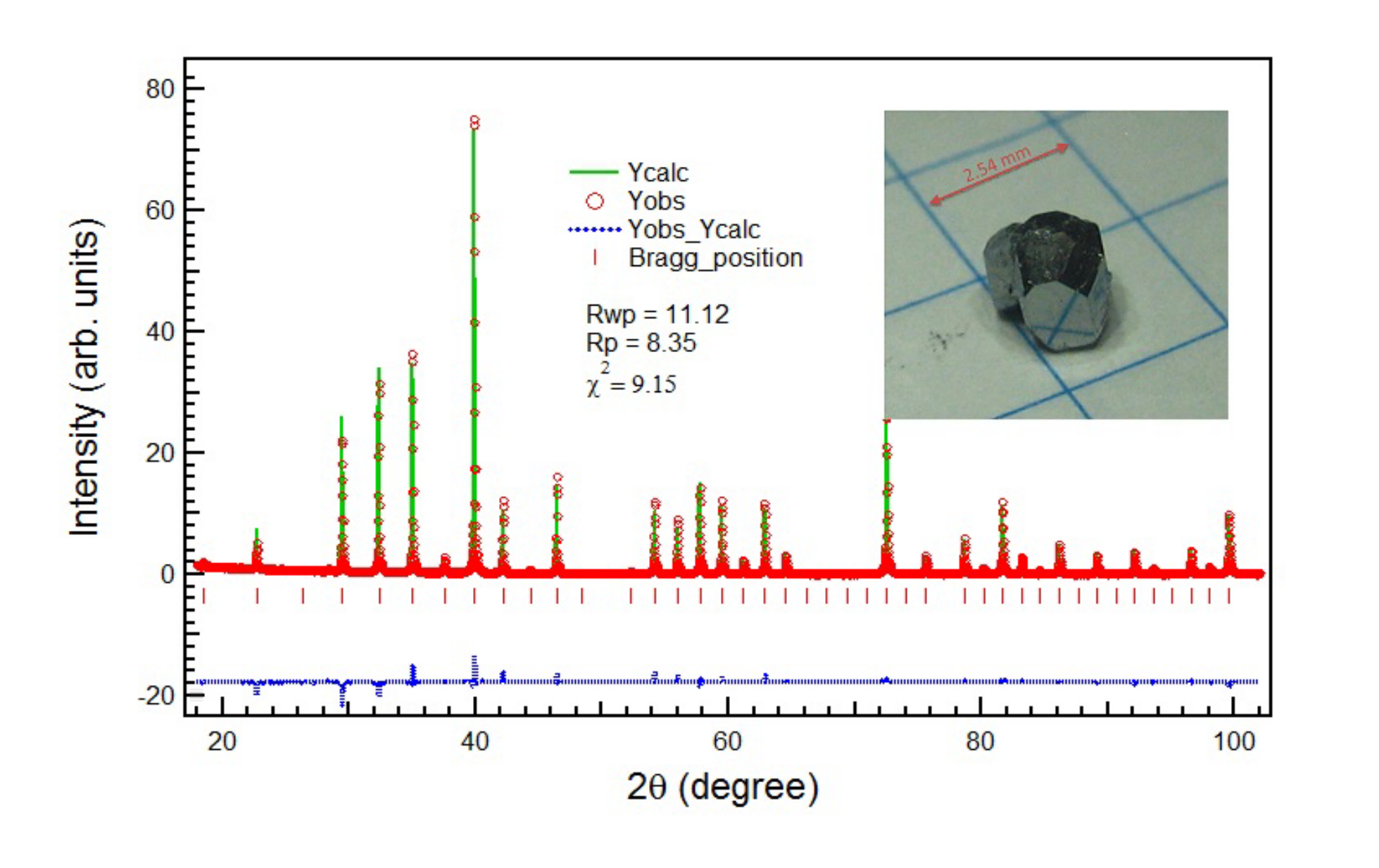}
\caption{(color online) Powder x-ray diffraction pattern of pulverized Mo$_3$Sb$_7$ single crystals recorded at room temperature. The observed (open circle) and
calculated (solid line) profiles are shown on the top. The vertical marks in the middle are
calculated positions of the Bragg peaks. The dashed line in bottom of the plot is the difference between calculated and observed intensities. Inset shows a piece of Mo$_3$Sb$_7$ single crystal.}
\label{PXRD-1}
\end{figure}

\section{Results and Discussions}

\subsection{Crystal growth and phase stability}

As-grown crystals are normally cube-like with truncated corners as shown in the inset of Fig.\,\ref{PXRD-1}. The typical dimension of a crystal is about 2-3\,mm. Elemental analysis confirmed that the crystals are stoichiometric Mo$_3$Sb$_7$ with no observable deviation. Figure\,1 shows the x-ray powder diffraction (XRD) data collected on ground Mo$_3$Sb$_7$ single crystals. The XRD data were analyzed by Rietveld refinement using the FullProf software and the Rietveld fit profiles are also presented in Fig.\,\ref{PXRD-1}. The XRD data indicate that the crushed crystals are single phase Mo$_3$Sb$_7$. Weak reflections from Sb were observed in some measurements. This small amount of Sb is from residual flux  on the surface of the crystals after the decanting and can be removed mechanically or by heating the crystals in evacuated and sealed quartz tube at 600$^o$C with the cold end at room temperature.

Growths with various charge/flux ratios were performed in order to optimize the growth and the yield. We noticed that raising the charge to flux ratio or cooling faster leads to smaller Mo$_3$Sb$_7$ crystals. No sizeable crystals were obtained using the above process once the Mo content is over 5\,at\%, even with a 24-hour homogenization at 1150$^o$C before cooling. From the presently available phase diagram,\cite{Mo-SbPD} Mo$_3$Sb$_7$ decomposes at 780$^o$C into (Mo) and liquid (Sb). However, a test growth stopped at 800$^o$C also yields Mo$_3$Sb$_7$ crystals from a Mo:Sb=1:49 melt. This suggests that there is a wide temperature and composition range where Mo$_3$Sb$_7$ and liquid (Sb) coexist. However, the liquidus temperature should  increase rapidly with increasing Mo content in the melt, which limits the charge/flux ratio during crystal growth. A Mo/Sb ratio of 1:49 has been found to reproducibly grow the largest crystals.

\begin{figure} \centering \includegraphics [width = 0.47\textwidth] {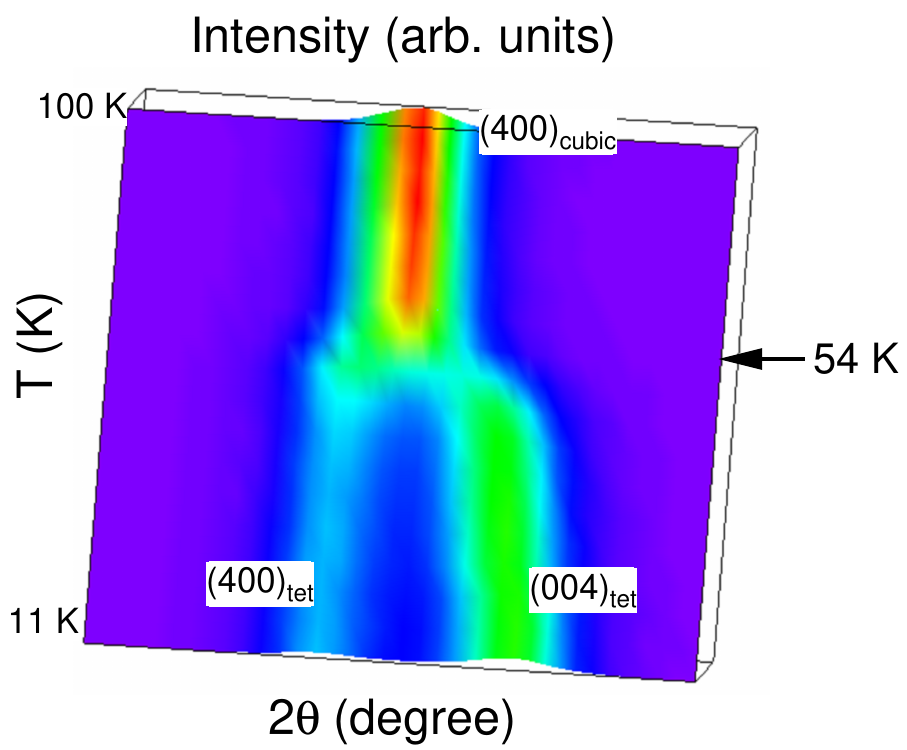}
\caption{(color online) Plot of the x-ray diffraction pattern for (400)$_{cubic}$ reflection in Mo$_3$Sb$_7$ as a function of temperature measured on a piece of oriented single crystal. The stronger intensity of (004)tet than that of (400)tet is due to the domain distribution. }
\label{SXRD-1}
\end{figure}

\begin{table*}[!ht]
\caption{Structure parameters for Mo$_3$Sb$_7$ with the space group \emph{Im}$\overline{3}$\emph{m} (no. 229) at room temperature. Atomic coordinates are : Mo (\emph{x},0,0), Sb1 (0.25,0,0.5), and Sb2 (\emph{x},\emph{x},\emph{x}). X-ray data were obtained from Rietveld refinement of powder diffraction pattern collected on pulverized single crystals. Neutron data were measured on a piece of single crystal of ~30\,mg at HB-3A. Lattice parameter a was determined to be 9.569(2)${\AA}$ by x-ray and 9.543(2)${\AA}$ by neutron diffraction. \\}
 \label{table0}
\begin{ruledtabular}
\begin{tabular} {lllllll}
 atom & site & \emph{x}(Xray/Neutron:X/N) &  \emph{U}$_{iso}$($\AA$$^2$)(X/N) \\
\hline
Mo & 12\emph{e}   & 0.3438(3)/0.34340(16)  &0.0060(3)/0.0041(5)     \\
Sb1 & 12\emph{d}   & 1/4 &  0.0084(3)/0.0068(5)        \\
Sb2& 16\emph{f}   & 0.16268(9)/0.16219(10) &0.0081(3)/0.0055(4)     \\
\end{tabular}
\end{ruledtabular}
\end{table*}

\begin{table*}[!ht]
\caption{Structure parameters for Mo$_3$Sb$_7$ with the space group \emph{I}4/\emph{mmm} (no. 139) at 4 K . Neutron data were measured on a piece of single crystal of ~30\,mg at HB-3A. The lattice parameters are a\,=\,9.551(4)${\AA}$, and c\,=\,9.523(2)${\AA}$ with a/c\,=\,1.003. R$_F$=0.0373, $\chi$$^2$\,=\,3.563 for 210 F$_o$$>$2$\sigma$(F$_o$).\\}
 \label{table0}
\begin{ruledtabular}
\begin{tabular} {lllllll}
 atom & site & \emph{x} & \emph{y} & \emph{z} &  \emph{U}$_{iso}$($\AA$$^2$) \\
\hline
 Mo1 & 8\emph{i}& 0.3436(2)   &  0     &  0   &   0.0016(6)  \\
 Mo2 & 4\emph{e} & 0     &  0     &  0.3442(4) &   0.0012(9)  \\
 Sb1 &4\emph{d} & 0     &  1/2     &  1/4   &   0.0013(11) \\
 Sb2 &8\emph{j} & 0.2501(6)   &  1/2     &  0   &   0.0012(6)  \\
 Sb3 &16\emph{m} & 0.16232(13) &  0.16232(13) &  0.1621(3) &   0.0010(4)  \\
\end{tabular}
\end{ruledtabular}
\end{table*}

\subsection{Structure}
The Rietveld refinement of the x-ray powder data shown in Fig.\,\ref{PXRD-1} confirmed that as-grown Mo$_3$Sb$_7$ single crystals crystallize in the Ir$_3$Ge$_7$ type structure with the cubic \emph{Im}$\overline{3}$\emph{m} (space group 229) symmetry at room temperature. The room temperature structure is further confirmed by a single crystal neutron diffraction measurement. Table 1 shows the crystallographic parameters refined for the room temperature x-ray powder diffraction pattern and the neutron single crystal diffraction data. The results agree well with each other and previous reports.\cite{StructureJP,StructureMRB}

A structural transition from cubic to tetragonal was reported to take place at 53\,K in polycrystalline samples. Single crystal x-ray and neutron diffraction were used to study this transition in our crystal. Figure \ref{SXRD-1} shows the evolution of the cubic (4\,0\,0) peak with temperature from 11\,K to 100\,K. Below 54\,K, the peak splits signaling the lowering of the symmetry. At 11 K, the a/c ratio was obtained to be 1.002, consistent with previous reports from powder x-ray diffraction.\cite{StructureJP,StructureMRB}

\begin{figure} \centering \includegraphics [width = 0.47\textwidth] {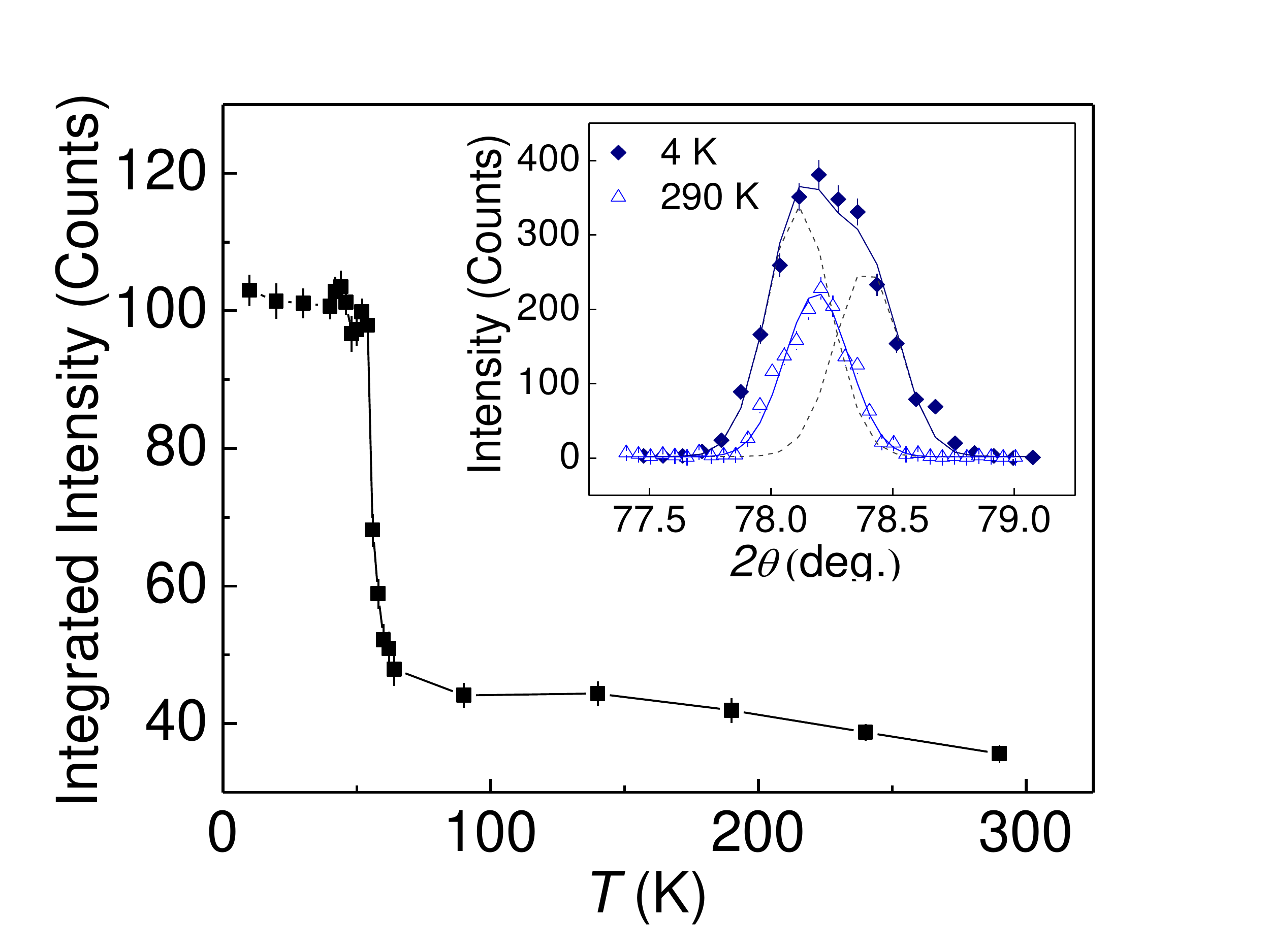}
\caption{(color online) Temperature dependence of the integrated intensity of the cubic (12\,0\,0) reflection measured by neutron single crystal diffraction on a Mo$_3$Sb$_7$ single crystal. Inset shows the peak splitting and the fitting of low temperature peaks. }
\label{Neutron-1}
\end{figure}

Figure \ref{Neutron-1} shows the temperature dependence of the integrated intensity of the cubic (12\,0\,0) nuclear peak measured with neutron single crystal diffraction. Two features are noteworthy: (1) the (12\,0\,0) broadens below T*\,=\,53\,K signaling the structure transition. Using the peak width of (12\,0\,0) fitted by Gaussian function at high temperature, the peak splitting was obtained by fitting with two Gaussian functions (see inset of Fig.\,\ref{Neutron-1}). (2) a rapid increase of the (12\,0\,0) peak intensity was observed upon cooling through the structural transition. At such a high q, the magnetic contribution to the intensity, if any, should be negligible. This peak intensity change is due to the weakened extinction associated with the cubic-tetragonal structural transition at low temperatures. Table 2 shows the crystallographic parameters refined for the neutron single crystal diffraction data collected at 4\,K. Refinement was performed with the space group \emph{I}4/\emph{mmm} (space group 139) and the crystal twinning was considered in the refinement. We noticed that the atomic displacement parameters (ADPs) at room temperature is 3-5 times larger than those at 4\,K. This change of ADPs also contributes to the intensity change of (12\,0\,0) peak; however, a large decrease in extinction in this crystal occurs at the structure transition. The larger ADPs at 300\,K may not be induced by the structure instabilities in the cubic phase. In Ru$_3$Sn$_7$ with the same structure but no structural instabilities, similar ADPs were also observed.\cite{Ru3Sn7Xtal}

\begin{figure} \centering \includegraphics [width = 0.47\textwidth] {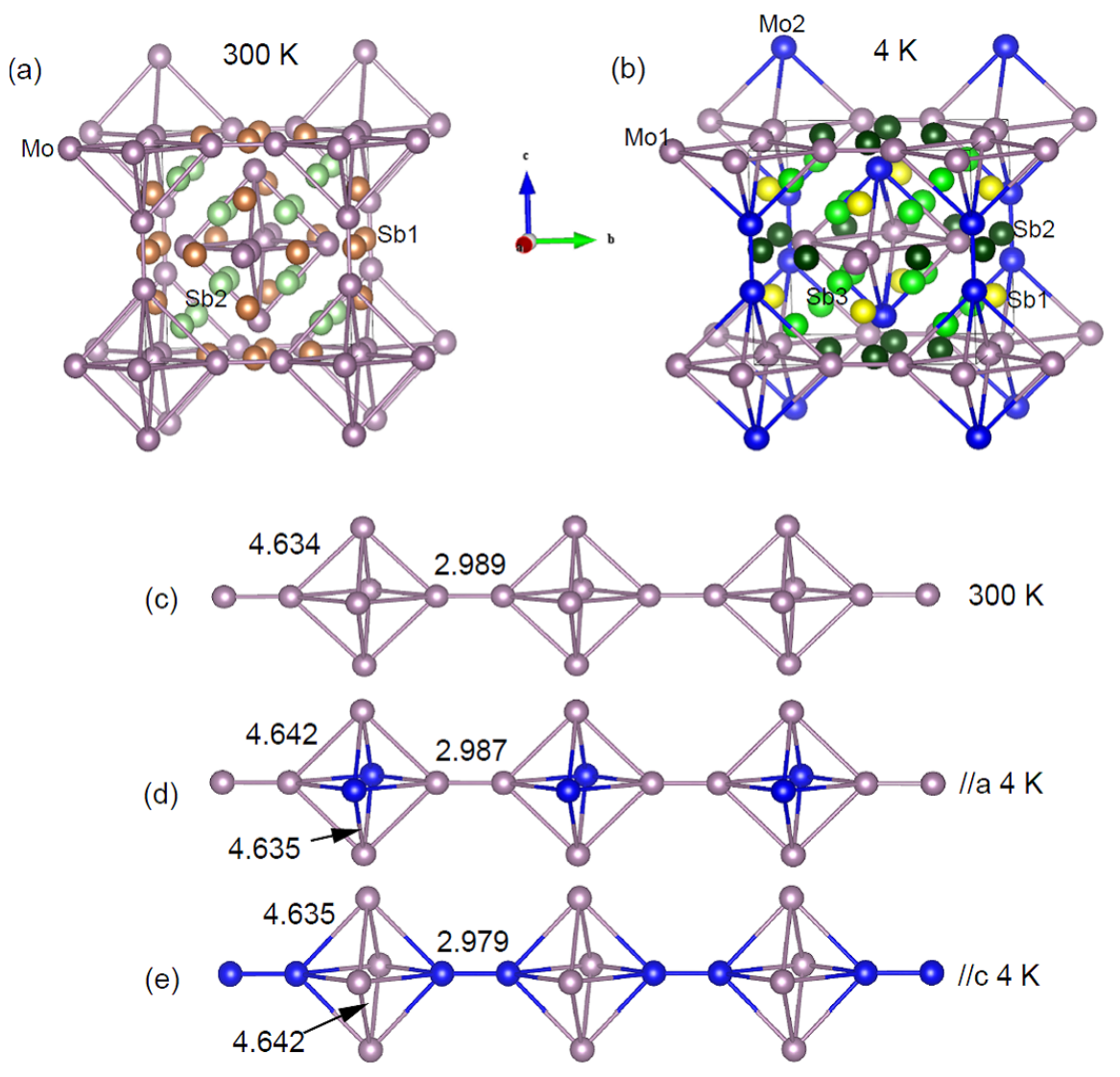}
\caption{(color online) Crystal structure of Mo$_3$Sb$_7$ at (a) room temperature, and (b) 4\,K. The -Mo-4Mo-Mo- structure unit and selected bond lengths (in unit of ${\AA}$) at (c) 300\,K, (d) 4\,K along a-axis, and (e) 4\,K along c-axis. }
\label{Structure1-1}
\end{figure}

Figure \ref{Structure1-1} shows the crystallographic structure with the parameters listed in Table 1 and Table 2. In the cubic phase above 53\,K, the Mo sublattice could be looked as three orthogonal -Mo-4Mo-Mo- chains as shown in Fig.\ref{Structure1-1} (a) and (c). In the tetragonal phase, Mo-Mo nearest neighbors along c-axis dimerize. As illustrated in Fig.\ref{Structure1-1} (d) and (e), the dimerization distinguishes the -Mo-4Mo-Mo- chains along c-axis from those along a- and b-axes.

\subsection{Superconducting transition}

\begin{figure} \centering \includegraphics [width = 0.47\textwidth] {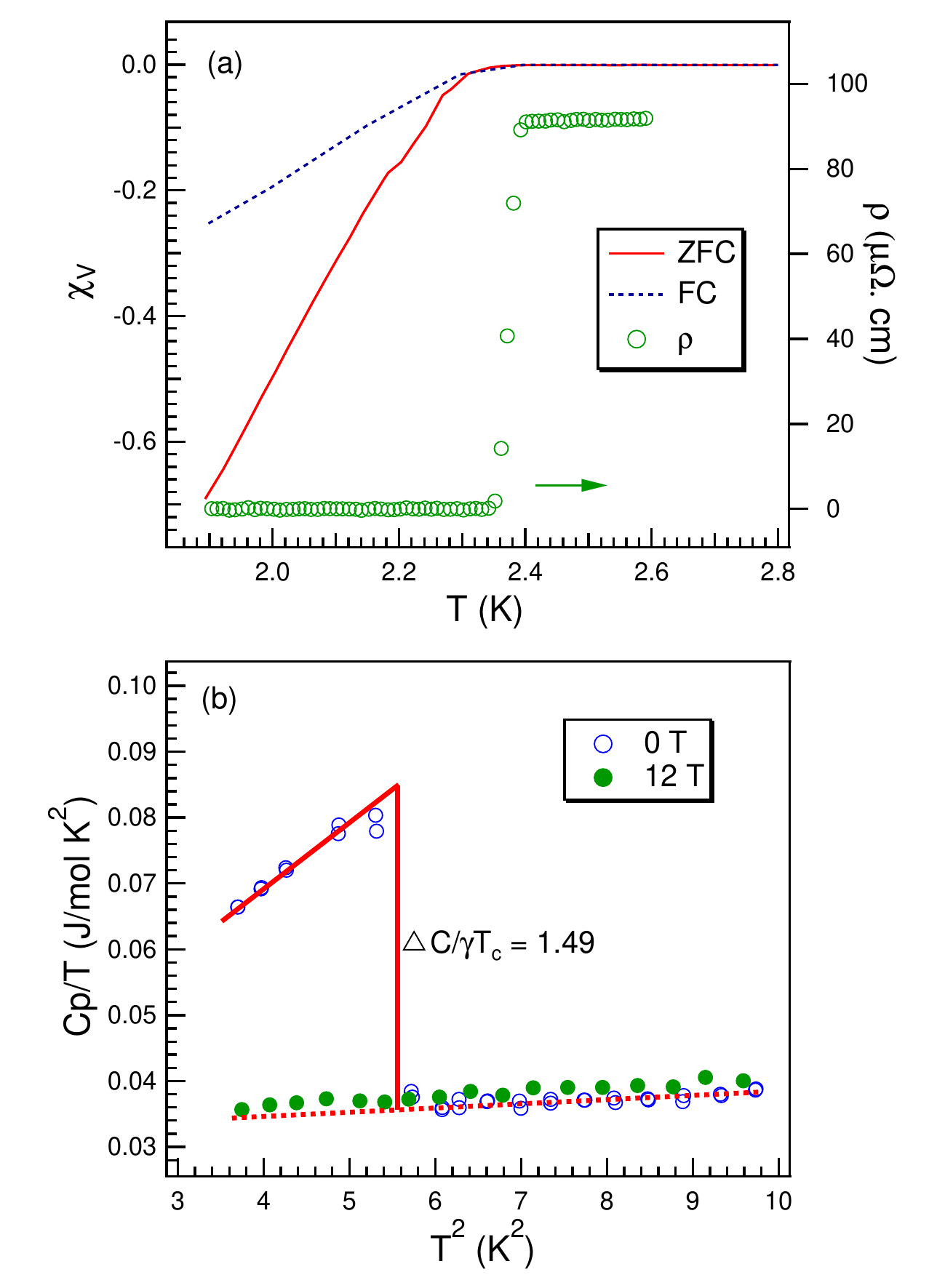}
\caption{(color online) (a) The temperature dependence of magnetic susceptibility (left) measured in both zero-field-cooling (ZFC) and field-cooling (FC) modes with a field of 10\,Oe and electrical resistivity (right). (b) The low-temperature specific heat of Mo$_3$Sb$_7$ single crystal divided by temperature as a function of temperature square measured under 0\,T (open circles) and 12\,T (solid circles) magnetic fields. The solid lines are guide to the eye. The dashed line is the linear fitting of the low temperature specific heat as described in the text.}
\label{Superconductivity-1}
\end{figure}

Superconductivity is observed to occur below T$_c$\,=\,2.35\,K, as indicated by the magnetization, electrical resistivity, and specific heat data shown in Fig.\,\ref{Superconductivity-1}.   Magnetization was measured in both zero-field-cooling (ZFC) and field-cooling (FC) modes with an applied magnetic field of 10\,Oe. The observed diamagnetism shown in Fig. \ref{Superconductivity-1}(a) confirms bulk superconductivity. The onset of the superconducting transition temperature is $\sim$2.35\,K, which agrees well with that determined from the temperature dependence of resistivity (also shown in Fig\,\ref{Superconductivity-1}(a)). Figure \ref{Superconductivity-1}(b) shows the low temperature specific heat data plotted as C$_p$/T vs. T$^2$. A specific heat jump at 2.35\,$\pm$\,0.05\,K was well resolved at the onset of superconductivity. High magnetic fields ($>$2\,T) suppress superconductivity and the specific heat jump disappears as shown in Fig.\,\ref{Superconductivity-1}(b). Literature values of T$_c$ vary from 2.02\,K to 2.30\,K, with the variation likely coming from sample quality.\cite{Dmitriev2007,Tran2008JOAM,Candolfi2009PRB,Candolfi2007PRL}  The current as-grown crystals posses the highest T$_c$ reported to date.

The fitting of specific heat data in the range 8\,$<$\,T$^2$\,$<$\,60\,K$^2$ to the standard power law, C$_p$\,=\,$\gamma$T\,$+$\,$\beta$T$^3$ yields $\gamma$\,=\,31.8(2)\,mJ/mol\,K$^2$ and $\beta$\,=\,0.64(1)\,mJ/mol\,K$^4$, where $\gamma$ is the Sommerfeld electronic specific heat coefficient and $\beta$ the coefficient of the Debye T$^3$ lattice heat capacity at low temperatures. The latter gives the Debye temperature $\theta$$_D$ with the following relation $\theta$$_D$\,=\,(12$\pi$$^4$N$_A$k$_B$n/5$\beta$)$^{1/3}$, where n is the number of atoms per formula unit, N$_A$ is Avogadro's constant and k$_B$ is Boltzmann's constant. With n\,=\,10 and $\beta$\,=\,0.64(1)\,mJ/mol\,K$^4$, the Debye temperature is $\theta$$_D$\,=\,248\,K for Mo$_3$Sb$_7$. We note that the Debye temperature is similar to that reported by Candolfi et al.,\cite{Candolfi2007PRL} but is smaller than that reported by Tran et al.\cite{Tran2008PRL,Tran2008Acta} The dimensionless specific heat jump $\triangle$C(T)/$\gamma$$_n$T$_c$ was determined to be 1.49, close to 1.56 reported by Tran et al.\cite{Tran2008Acta} This value is slightly larger than the BCS value of 1.43 for the weak-coupling limit.

The density of states at the Fermi level D(E$_F$) can be obtained from the Sommerfeld coefficient $\gamma$ according to:
\begin{equation}
\gamma =\frac{\pi ^{2 }k_{B}^{2}}{3}D(E_{F})(1+\lambda _{e-ph})
\end{equation}

\noindent where $\lambda$$_{e-ph}$ is the electron-phonon coupling constant and could be estimated with the McMillan equation:
\begin{equation}
\lambda _{e-ph} = \frac{1.04+\mu ^*ln(\theta_D/1.45T_c)}{(1-0.62\mu^*)ln(\theta_D/1.45T_c)-1.04}
\end{equation}

\noindent where $\mu$$^*$ is the Coulomb pseudopotential and usually taken between 0.1 and 0.15. With $\mu$$^*$\,=\,0.15, $\Theta$$_D$\,=\,248\,K, and T$_c$\,=\,2.35\,K, $\lambda$$_{e-ph}$ is 0.59 consistent with a previous report.\cite{Tran2008Acta} The  density of states at the Fermi level D(E$_F$) is calculated to be 8.5\,states/eV\,f.u., where f.u. means formula unit. The value is comparable to that obtained by theoretical calculations. \cite{DFTcal2011}

\begin{figure} \centering \includegraphics [width = 0.47\textwidth] {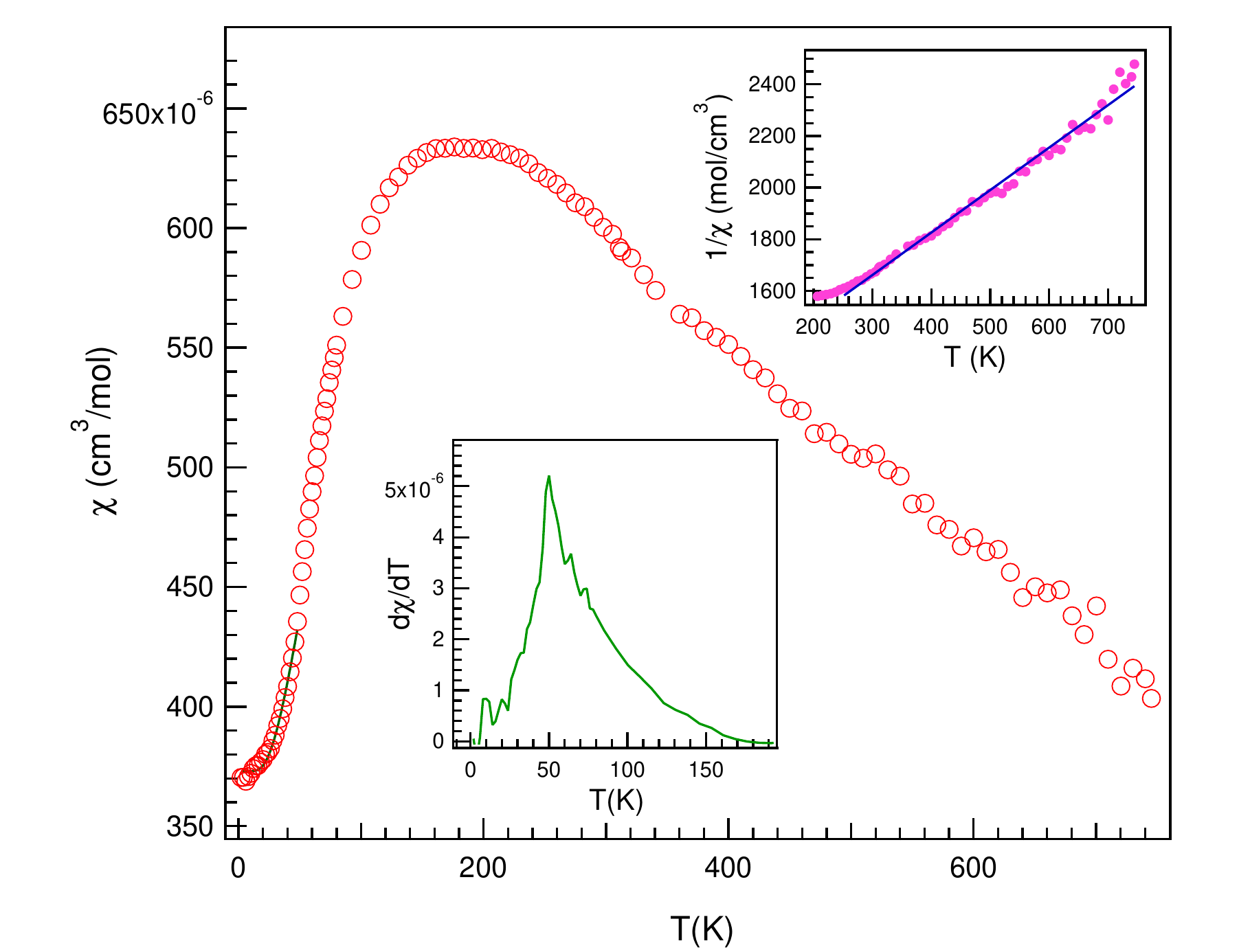}
\caption{(color online) The temperature dependence of magnetic susceptibility of Mo$_3$Sb$_7$ single crystal measured under a magnetic field of 60\,kOe in the temperature range 2\,K\,$\leq$\,T\,$\leq$\,750\,K. The solid curve below 50\,K shows the fitting as described in text. The lower inset shows the temperature derivative of magnetic susceptibility, d$\chi$(T)/dT. The upper inset shows the linear fitting of $\chi$(T)$^{-1}$.}
\label{Mag-1}
\end{figure}

\subsection{Physical properties in normal state}

Figure\,\ref{Mag-1} shows the temperature dependence of the magnetic susceptibility, $\chi(T)$, measured with an applied field of 60\,kOe in the temperature range 2\,K\,$\leq$\,T\,$\leq$\,750\,K. For $\chi(T)$ measured above room temperature, the contribution from the sample holder was appropriately corrected. With decreasing temperature, $\chi(T)$ increases and shows a broad maximum around T$_{max}$\,=\,175\,K which is followed by a sharp drop around 50\,K. The sharp drop is best illustrated by the derivative d$\chi$$/$dT, which shows a distinct maximum around 50\,K as shown in inset of Fig.\,5. The features below room temperature agree with previous reports.\cite{Candolfi2007PRL,Tran2008PRL} It's noteworthy that below 20\,K $\chi(T)$ shows negligible temperature dependence. While in all previous reports on the magnetic properties of polycrystalline samples, a Curie-Weiss (CW)-like tail is observed below 50\,K, which has been attributed to unpaired spins. The absence of the CW-like tail further confirms the high quality of our crystals. The isothermal magnetization (not shown) measured at 5\,K and room temperature indicates no ferromagnetic component.

As shown in Fig.\,\ref{Mag-1}, $\chi(T)$ below 50 K was fit with a spin gap function $\chi$(T)\,=\,$\chi$(0)\,+\,\emph{b}exp(-$\Delta$/k$_B$T), where $\chi$(0) is a temperature independent term, \emph{b} is a constant, $\Delta$ is the spin gap, k$_B$ is the Boltzmann constant. The best fitting was obtained with the following parameters: $\chi$(0)\,=\,3.7$\times$10$^{-4}$\,cm$^3$/mol, \emph{b}\,=\,5.8\,$\times$10$^{-4}$cm$^3$/mol, and $\Delta$/k$_B$\,=\,110(6) K. The gap is similar to that reported  on polycrystalline samples.\cite{Tran2008PRL}

The magnetic susceptibility of Mo$_3$Sb$_7$ in the temperature interval 220\,K\,$\leq$\,T\,$\leq$\,300\,K has been reported\cite{Tran2008PRL} to follow Curie-Weiss law. The upper inset of Fig.\,\ref{Mag-1} shows the linear fitting of $\chi(T)$$^{-1}$ above 300\,K. The fitting yields  $\theta$\,=\,-1078\,K and $\mu$$_{eff}$\,=\,1.67\,$\mu$$_{B}$/Mo. The effective magnetic moment is close to the expected value for Mo ions with S=1/2. While such a large negative Weiss constant suggests strong antiferromagnetic interaction between localized Mo moments.

\begin{figure} \centering \includegraphics [width = 0.47\textwidth] {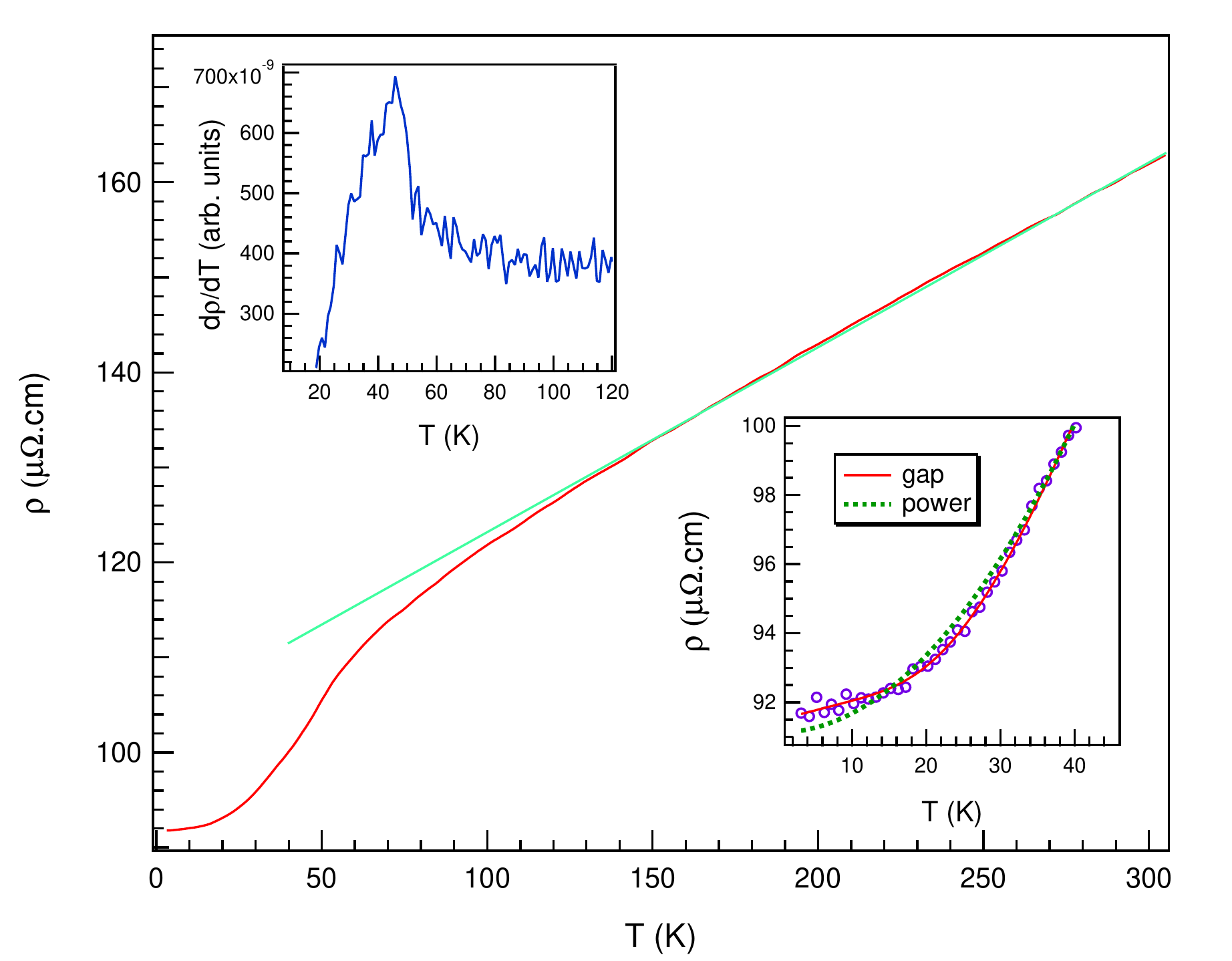}
\caption{(color online) The temperature dependence of electrical resistivity of Mo$_3$Sb$_7$ single crystal. The solid line highlights the linear temperature dependence of electrical resistivity at high temperatures. The upper inset shows the derivative of electrical resistivity, d$\rho$/dT. The lower inset compares the fitting with power law and a gap function as described in text.}
\label{RT-1}
\end{figure}

Figure \ref{RT-1} shows the temperature dependence of electrical resistivity in the temperature range 3\,K\,$\sim$\,300\,K. The electrical resistivity at room temperature is about 160\,$\mu$$\Omega$.cm. As highlighted by the solid line in the Figure, at temperatures above $\sim$150\,K, the electrical resistivity  shows a linear temperature dependence consistent with the report by Bukowski et al.\cite{Bukowski2002SSC} Starting around 50\,K, the electrical resistivity shows a sharp drop. This sharp drop can be better illustrated by the derivative of the electrical resistivity shown in the upper inset. At even lower temperatures, the resistivity levels off with cooling and finally drops to zero at T$_c$. The low temperature electrical resistivity in polycrystalline samples has been found to fit a $\rho$$($T$)$\,=\,$\rho$$_0$\,$+ \,$\emph{b}T$^\emph{n}$ law with \emph{n}\,=\,2.\cite{Candolfi2007PRL} The dashed curve in the lower inset shows the fitting, and obviously the fit is not acceptable for the data below $\sim$40\,K. A much better fit can be obtained with n around 2.4, which has been observed when fitting $\rho$$($T$)$ measured on crystals from different batches. We also tried to fit the low temperature electrical resistivity with a gap function $\rho$$($T$)$\,=\,$\rho$$_0$\,$+$\,\emph{c}T\,$+$\,\emph{d}exp$($-$\Delta$$\rho$/K$_B$T) which has been used to argue against the spin fluctuation scenario and point to a gap feature.\cite{Tran2008PRL} As shown in the lower inset of Fig.\,\ref{RT-1}, a good fit could be obtained with $\rho$$_0$\,=\,91.5\,$\mu$$\Omega$.cm, \emph{c}\,=\,5.58\,$\times$10$^{-8}$$\Omega$.cm.K$^{-1}$, \emph{d}\,=\,9.36\,$\times$10$^{-5}$$\Omega$.cm, and $\Delta$$\rho$/K$_B$\,=\,107(2)\,K.

The superconducting pairing mechanism at low temperatures is still under hot debate despite much effort. Nazir et al. compared the electronic properties of the cubic and tetragonal phase in  Mo$_3$Sb$_7$ by density functional theory calculations.\cite{DFTcal2011} A higher density of state at Fermi level is found in the tetragonal phase which has a higher propensity to magnetism than the cubic phase. Spin fluctuations were argued to coexist with superconductivity in Mo$_3$Sb$_7$.\cite{Candolfi2007PRL} One important evidence for this argument is the quadratic temperature dependence of electrical resistivity and magnetic susceptibility at low temperatures in the tetragonal phase. The absence of the expected quadratic temperature dependence signals that the electron-paramagnon interaction should be moderate.\cite{Oles2008}

X-ray diffraction on oriented single crystals (Fig. 2) and neutron single crystal diffraction (Fig 3) clearly show the symmetry lowering from cubic to tetragonal while cooling. The shortened Mo-Mo bond length along the crystallographic c-axis signals only 1/3 Mo-Mo pairs dimerize. The fact that spin singlet state is only a fraction of the magnetic ground state explains the large magnetization at low temperatures.  On the other hand, the above feature distinguishes Mo$_3$Sb$_7$ from other low temperature superconductors by the coexistence of spin dimer singlet and superconductivivty. In a spin dimer system, spin fluctuations from the triplet excitations could induce an attractive interaction between conduction electrons leading to the superconducting pairing.\cite{Matsumoto2005} An preliminary study on Te or Ru doped Mo$_3$Sb$_7$ suggests that a few percent of Te or Ru substitution eliminates the structure transition and superconductivity survives in the cubic phase. Thus the structure transition or possible spin fluctuations from the triplet excitation cannot be the only driving force for the superconducting pairing.

\begin{figure} \centering \includegraphics [width = 0.47\textwidth] {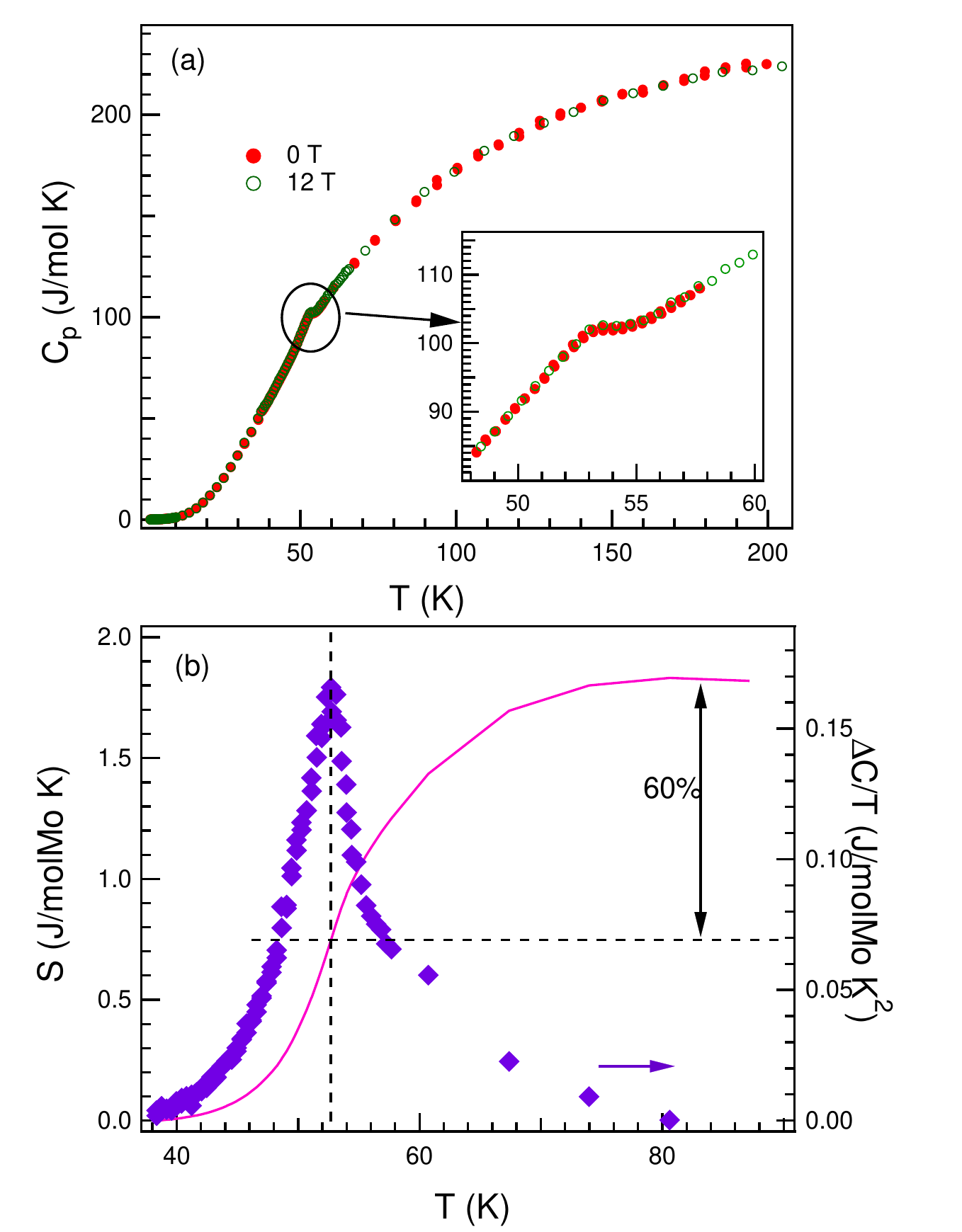}
\caption{(color online) (a) The temperature dependence of specific heat of Mo$_3$Sb$_7$ single crystal measured under 0 and 12 T magnetic fields. The inset highlights the weak lambda anomaly around 53\,K. (b) C$_p$/T versus T and the numerical integration of the entropy S after subtracting the lattice specific heat using Ru$_3$Sn$_7$ as phonon reference. }
\label{Cp-1}
\end{figure}

Figure \ref{Cp-1} shows the temperature dependence of the specific heat, C$_p$(T), measured under 0\,T and 12\,T applied magnetic fields. The magnetic field of 12\,T suppresses superconductivity and the specific heat jump at T$_c$ disappears and the details were shown in Fig.\,\ref{Superconductivity-1}(b). The heat capacity at 200\,K attains a value of ~226\,J/mol K, which is a little smaller than the classical high temperature Dulong-Petit value of 3nR\,=\,249\,J/mol K at constant volume, where R is the molar gas constant and n\,=\,10 is the number of atoms per formula unit. This agrees with the determined Debye temperature, which is slightly below room temperature. The inset of Fig.\,\ref{Cp-1} highlights the details around 50\,K. A weak lambda anomaly is well resolved around T\,=\,53\,K. No field dependence was observed for this anomaly with magnetic fields up to 12\,T.

In order to estimate the entropy change around 53\,K, we subtracted the lattice contribution to the specific heat using Ru$_3$Sn$_7$ as a reference.\cite{Ru3Sb7} Figure \ref{Cp-1}(b) shows the residual specific heat plotted as $\Delta$C$_p$/T vs T. The entropy change, which is calculated as $\int$$\Delta$C$_p$/T dT, is 1.80\,J/molMo\,K and about 1/3 of the expected value. This suggests that the entropy change is from Mo spins. As illustrated in Fig.\,\ref{Cp-1}(b), about 60\,$\%$ entropy change takes place above 53\,K.

\begin{figure} \centering \includegraphics [width = 0.47\textwidth] {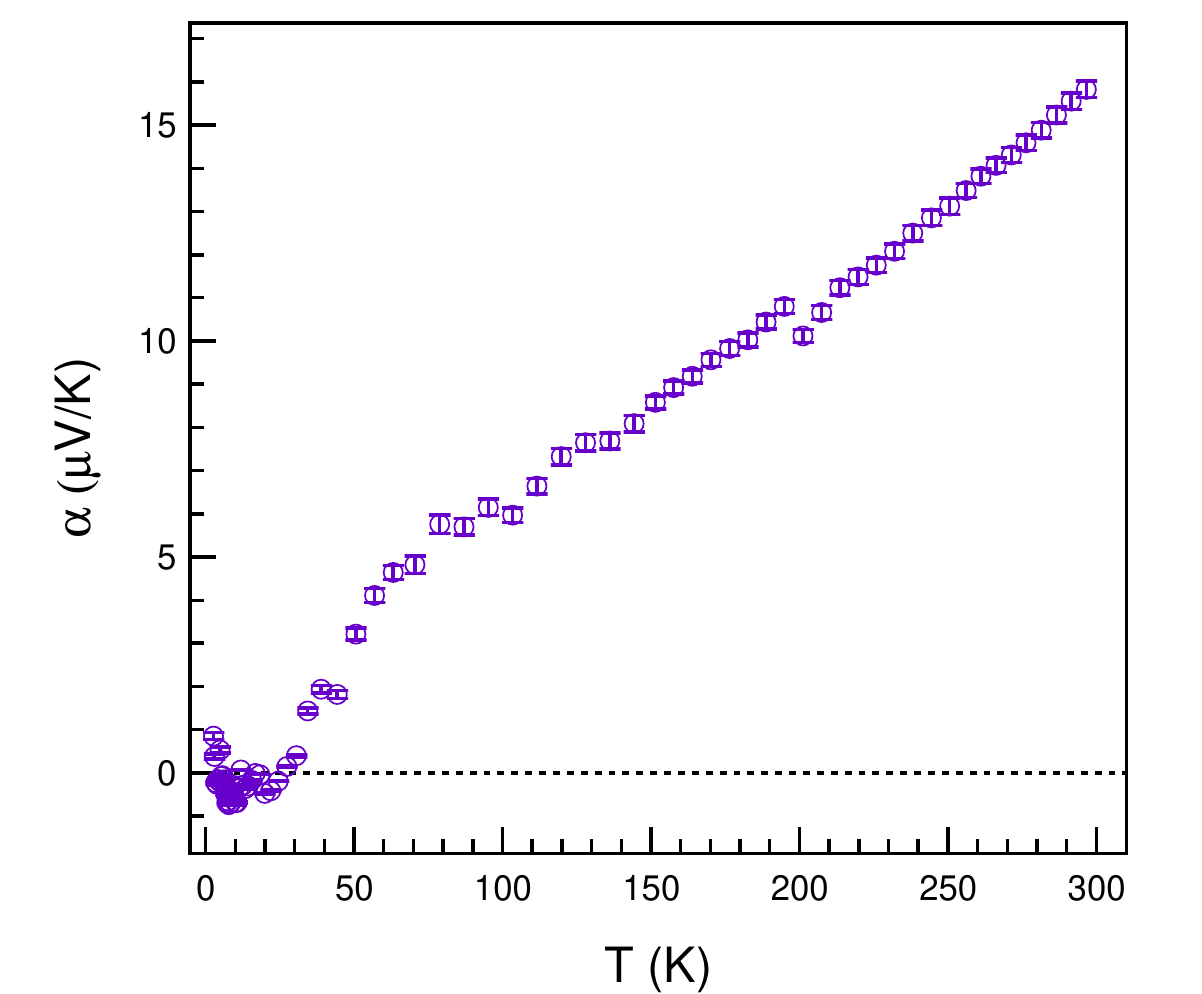}
\caption{(color online) The temperature dependence of thermopower of Mo$_3$Sb$_7$ single crystal. The dashed line is zero line.}
\label{Seeb-1}
\end{figure}

Figure \ref{Seeb-1} shows the evolution of the thermopower, $\alpha$(T), with temperature. A room temperature value of $\sim$16$\mu$V/K agrees with the measurement on a polycrystalline sample.\cite{Candolfi2009PRB} Above 100\,K, $\alpha$(T) shows a nearly linear temperature dependence. Below about 60\,K, $\alpha$(T) decreases quickly when cooling; this sharp drop resembles that in the temperature dependence of electrical resistivity around 53\,K. Below 30\,K, $\alpha$(T) stays around zero. It's interesting to note that $\rho$(T) levels off and is nearly temperature independent below 30\,K. The thermopower is proportional to the logarithmic derivative of the density of states (DOS) with respective to energy at the Fermi level, while the electrical conductivity increases with DOS. The observed change of $\alpha$(T) and $\rho$(T) suggests an electronic structure change across the structure transition.

\begin{figure} \centering \includegraphics [width = 0.47\textwidth] {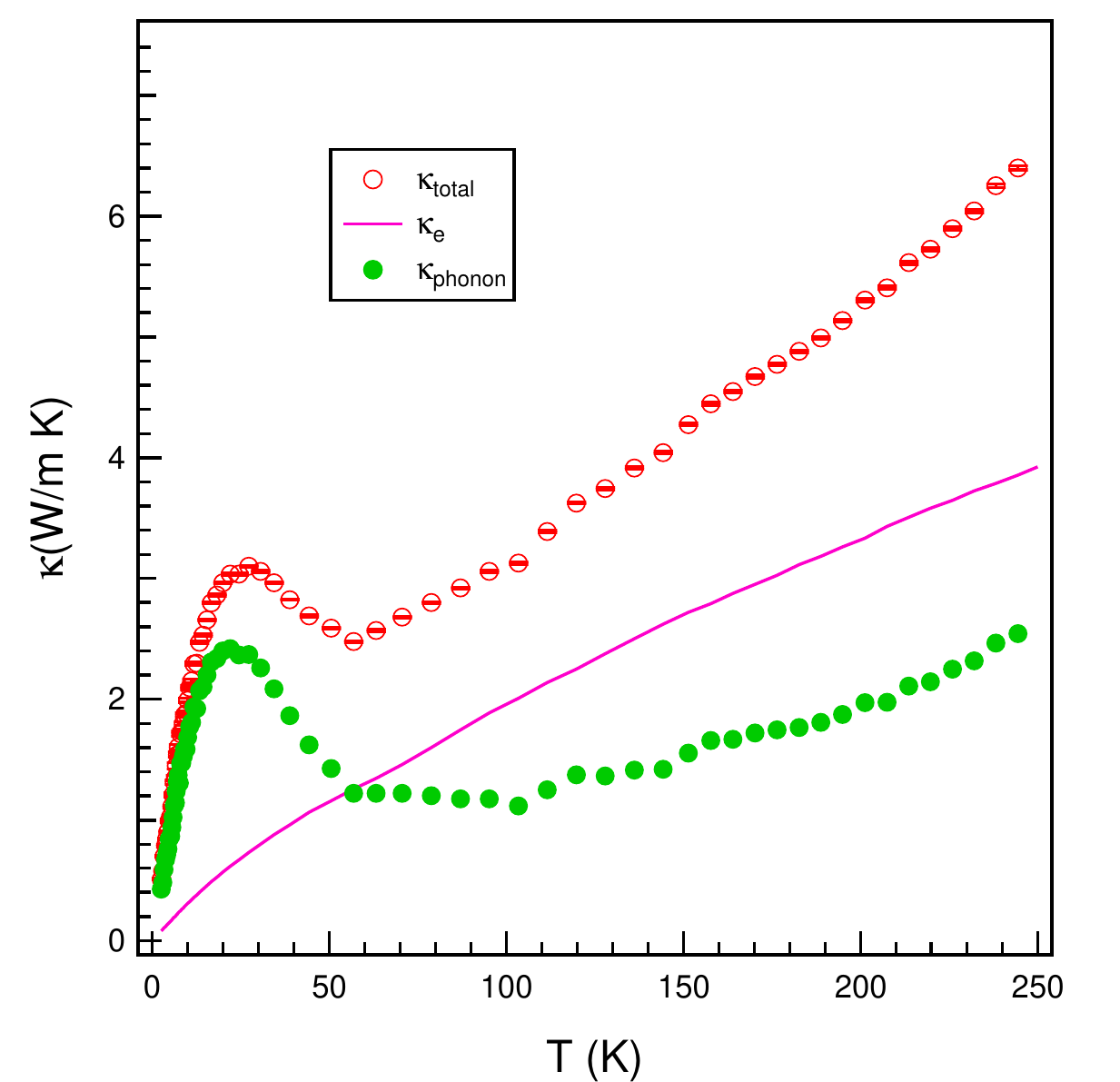}
\caption{(color online) The temperature dependence of thermal conductivity of Mo$_3$Sb$_7$ single crystal. See the text for the calculation of electron and phonon thermal conductivities.}
\label{Kappa-1}
\end{figure}

Figure \ref{Kappa-1} shows the temperature dependence of the thermal conductivity in the normal state. With decreasing temperature, the thermal conductivity decreases from room temperature down to ~50\,K; below 50\,K, a maximum near 20\,K was observed. Since no long range magnetic order was observed in Mo$_3$Sb$_7$, possible heat carriers are electrons and phonons. Thus the total thermal conductivity can be described as:
$\kappa$$_{total}$\,=\,$\kappa$$_{e}$\,+\,$\kappa$$_{p}$, where $\kappa$$_{e}$ is the electronic thermal conductivity, and  $\kappa$$_{p}$  the lattice thermal conductivity. $\kappa$$_{e}$ was estimated from the electrical resistivity data using the Wiedemann-Franz law: $\kappa$$_{e}$=\,LT/$\rho$, where L is the Lorenz constant taken to be equal to 2.44\,$\times$\,10$^{-8}$\,V$^2$/K$^2$, T is the absolute temperature, and $\rho$ is the electrical resistivity. Figure 9 shows the temperature dependence of $\kappa$$_{e}$ and $\kappa$$_{p}$; the latter was obtained by subtracting $\kappa$$_{e}$  from the total thermal conductivity. $\kappa$$_{e}$  decreases while cooling in the whole temperature range studied. We noticed that $\kappa$$_{p}$ is rather low in the whole temperature range. This might be due to large number of heavy atoms in the primitive unit cell or strong electron-phonon scattering. The maximum near 20\,K for $\kappa$$_{p}$  is typical for a crystalline material though the maximum occurs at a low value. The striking feature for $\kappa$$_{p}$ is a glassy behavior above 53\,K where the structure transition takes place.  This kind of phonon glass behavior has been observed in a wide variety of materials, such as RVO$_3$,\cite{Yan2004PRL} RCoO$_3$,\cite{YanBLF} La$_4$Ru$_2$O$_{10}$,\cite{Fran2011} and filled skutterudite antimonides.\cite{Sales1997} In the oxides with localized electrons, bond-length fluctuations induced by spin and/or orbital fluctuations disturb the phonon thermal conductivity through spin-orbital-lattice coupling. While in filled skutterudite antimonides, the rattling mode seriously damps the lattice vibration. We note that neutron single crystal diffraction didn't observe abnormal atomic displacement parameters for either Mo or Sb in Mo$_3$Sb$_7$.\cite{SalesADP} Thus, the phonon glass behavior of Mo$_3$Sb$_7$ might signal the effect of the structural instability resulting from the competing NN and NNN magnetic interactions. The cubic-tetragonal structural transition relieves the magnetic frustration thereby recovering the phonon heat transport in the tetragonal phase. This might be further enhanced by the reduced spin scattering due to the accompanied spin gap opening.

\section{Summary}

Mo$_3$Sb$_7$ single crystals were successfully grown out of Sb flux. Our growth study suggests that there is a wide temperature and composition range where Mo$_3$Sb$_7$ and liquid (Sb) coexist. The sizeable crystals enabled the first neutron single crystal diffraction study of the high temperature cubic phase and the low temperature tetragonal phase. The physical properties are generally in line with previous reports on polycrystalline samples. However, the following features are noteworthy: (1) Magnetization, electrical resistivity and specific heat measurements observed superconductivity at 2.35\,$\pm$\,0.05\,K in the as-grown crystals, which is the highest T$_c$ reported to date for Mo$_3$Sb$_7$. The corresponding dimensionless specific heat jump $\triangle$C(T)/$\gamma$$_n$T$_c$ was determined to be 1.49, which is slightly larger than the BCS value of 1.43 for the weak-coupling limit. (2) A weak lambda anomaly was observed at 53\,K in the temperature dependence of specific heat and no field effect was observed with magnetic field up to 12\,T. The entropy change across the transition is estimated to be 1.80\,J/molMo\,K and 60\% entropy change takes place above 53\,K. (3) The magnetic susceptibility shows negligible temperature dependence below $\sim$20\,K suggesting the crystals are free of magnetic impurities. Above room temperature, the magnetic susceptibility follows CW law and a moment of 1.67\,$\mu$$_B$/Mo is close to the expected value for Mo ions with spin S\,=\,1/2. Thus the electronic state of Mo has both localized and itinerant characters. (4) The electrical resistivity below 50\,K doesn't show a quadratic temperature dependence. Analysis of the transport and magnetization data suggests the opening of a spin gap of 110\,K accompanying the structure transition. (5) A phonon glass behavior was observed above 53\,K suggesting strong scattering of phonons by lattice instabilities. The abrupt change around 53\,K in the temperature dependence of electrical resistivity and thermopower suggests an electronic structure change across the structural transition.

The sizeable high quality single crystals allow studies of intrinsic properties of Mo$_3$Sb$_7$. Moreover, preliminary growths have demonstrated that crystals with various substitutional dopants, for either Mo or Sb, can be grown with controlled doping using the reported growth protocol. This enables various future studies that will explore the relationship between the structure instability, magnetism and superconductivity, and the origin of the promising thermoelectric performance of doped Mo$_3$Sb$_7$.

\section{Acknowledgments}
The authors thank D. Parker for helpful discussion. Research was supported by the U.S. Department of Energy, Basic Energy Sciences, Materials Sciences and Engineering Division. Work at ORNL's High Flux Isotope Reactor was sponsored by the Scientific User Facilities Division, Office of Basic Energy Sciences, U.S. Department of Energy.


\begin{references}

\bibitem{Bukowski2002SSC}Z. Bukowski, D. Badurski, J. Stepien-Damm, and R. Troc, Solid State Communication \textbf{123}, 282 (2002).

\bibitem{Candolfi2007PRL}C. Candolfi, B. Lenoir, A. Dauscher, C. Bellouard, J. Hejtmanek, E. Santava, and J. Tobola, Phys. Rev. Lett \textbf{99}, 037006 (2007).

\bibitem{Tran2008PRL}V. H. Tran, W. Muller, and Z. Bukowski, Phys. Rev. Lett. \textbf{100}, 137004 (2008).

\bibitem{Tran2009JP}V. H. Tran, A. D. Hillier, D. TAdroja, Z. Bukowski, and W. Miiller, J. Phys.: Condens. Matter \textbf{21}, 485701 (2009).

\bibitem{Oles2008}B. Wiendlocha, J. Tobola, M. Sternik, S. Kaprzyk, K. Parlinski, and A. M. Oles, Phys. Rev. B \textbf{78}, 060507(R) (2008).

\bibitem{Parker2011}D. Parker, M. Du, and D. J. Singh, Phys. Rev. B \textbf{83}, 245111 (2011).

\bibitem{HTTEMo3SbTe}C. Candolfi, B. Lenoir, C. Chubilleau, A. Dauscher, and E. Guilmeau,J. Phys.: Condens. Matter \textbf{22}, 025801 (2010).

\bibitem{HTTERu}C. Candolfi, J. Leszczynski, P. Masschelein, C. Chubilleau, B. Lenoir, A. Dauscher, E. Guilmeau, J. Hejtmanek, S. J. Clarke, R. I. Smith, J. Electronic Mater., \textbf{39}, 2132 (2010).

\bibitem{Jeff2011}Xiaoya Shi, Yanzhong Pei, G. Jeffrey Snyder, and Lidong Chen, Energy \& Environmental Science, \textbf{4}, 4086 (2011).

\bibitem{ValenceBondXtal}T. Koyama, H. Yamashita, Y. Takahashi, T. Kohara, I. Watanabe, Y. Tabata, and H. Nakamura, Phys. Rev. Lett. \textbf{101}, 126404 (2008).

\bibitem{StructureJP}H. Okabe, S. Yano, T. Muranaka, and J. Akimitsu, J. Phys.:Conference Series, \textbf{150}, 052196 (2009).

\bibitem{StructureMRB}T. Koyamaa, H. Yamashitaa, T. Koharaa, Y. Tabatab, H. Nakamurab, Materials Research Bulletin \textbf{44}, 1132 (2009).

\bibitem{Jensen1966}P. Jensen, and A. Kjekshus, Acta Chemica Scandinavica \textbf{20}, 417 (1966).

\bibitem{Dmitriev2007}V. M. Dmitriev, L. F. Rybaltchenko, E. V. Khristenko, L. A. Ishchenko, Z. Bukowski, and R. Troc, Low Temp. Phys. \textbf{33}, 295 (2007).

\bibitem{Tran2008JOAM}V. H. Tran, E. Bauer, A. Galatanu, and Z. Bukowski, J. Optoelectronics and Advanced Materials \textbf{10}, 1630 (2008).

\bibitem{FullProf}J Rodrguez-Carvajal, Physica B \textbf{192}, 55 (1993).

\bibitem{HB3A}B. C. Chakoumakos, H. Cao, F. Ye, A. D. Stoica, M. Poovici, M. Sundaram, W. Zhou, J. S. Kicks, G. W. Lynn, and R. A. Riedel, J. Appl. Crystallogr. \textbf{44}, 655 (2011).

\bibitem{Mo-SbPD}Brewer L., and Lamoreaux R.H., Mo-Sb (Molybdenum-Antimony), Binary Alloy Phase Diagrams, II Ed., Ed. T.B. Massalski, \textbf{3}, 2660 (1990).


 \bibitem{Ru3Sn7Xtal}B. C. Chakoumakos, and D. Mandrus, J. Alloy Compd \textbf{281}, 157 (1998).


\bibitem{Candolfi2009PRB}C. Candolfi, B. Lenoir, A. Dauscher, E. Guilmeau, J. Hejtmanek, J. Tobola, B. Wiendlocha, and S. Kaprzyk, Phys. Rev. B \textbf{79}, 035114 (2009).

\bibitem{Tran2008Acta}V. H. Tran, W. Miller, and Z. Bukowski, Acta Materiallia \textbf{56}, 5694 (2008).

\bibitem{DFTcal2011}S. Nazir, S. Auluck, J. J. Pulikkotil, N. Singh, and U. Schwingenschlogl, Chem. Phys. Lett. \textbf{504}, 148 (2011).


\bibitem{Matsumoto2005}M. Matsumoto, and M. Koga, Progress of Theoretical Physics Supplement \textbf{159}, 371 (2005).

\bibitem{Ru3Sb7}V. H. Tran, and W. Miller, Acta Physica Polonica A, \textbf{115}, 83 (2009).

\bibitem{Yan2004PRL}J.-Q. Yan, J.-S. Zhou, and J. B. Goodenough, Phys. Rev. Lett. \textbf{93}, 235901 (2004).

\bibitem{YanBLF}J.-Q. Yan, J.-S. Zhou, and J. B. Goodenough, Phys. Rev. B \textbf{69}, 134409 (2004).

\bibitem{Fran2011}B. Rivas-Murias, H. D. Zhou, J. Rivas, and F. Rivadulla, Phys. Rev. B \textbf{83}, 165131 (2011).

\bibitem{Sales1997}B. C. Sales, D. Mandrus, B. C. Chakoumakos, V. Keppens, and J. R. Thompson, Phys. Rev. B \textbf{56}, 15081 (1997).

\bibitem{SalesADP}B. C. Sales, D. Mandrus, B. C. Chakoumakos, Semiconductors and Semimetals \textbf{70}, 1 (2001).

\end{references}
\end{document}